\documentclass[]{aastex631}

\usepackage{amsmath}
\usepackage{epstopdf} 
\usepackage{graphicx}
\usepackage{epsfig}

\begin{document}

\title{Variations in Dominant Wave Period in the Solar Atmosphere}

\author[0000-0002-0786-7307]{Pradeep Kayshap}
\affiliation{School of Advanced Sciences and Languages, VIT Bhopal University, Kothrikalan, Sehore Madhya Pradesh - 46611}

\author{K. Murawski}
\affiliation{Institute of Physics,
            University of M. Curie-Sk{\l}odowska, 
             Pl.\ M.\ Curie-Sk{\l}odowskiej 1, 
             PL-20-031 Lublin, Poland}

\author{Z. E. Musielak}
\affiliation{Department of Physics, University of Texas at Arlington, Arlington, TX 76019, USA}

\author{Suresh Babu}
\affiliation{School of Advanced Sciences and Languages, VIT Bhopal University, Kothrikalan, Sehore Madhya Pradesh - 46611}

%\author{Blazej Ku{\'z}ma}
%\affiliation{Shenzhen Key Laboratory of Numerical Prediction for Space Storm,
%Harbin Institute of Technology, Shenzhen 51805, People’s Republic
%of China}

%\author{Ding Yuan}
%\affiliation{Shenzhen Key Laboratory of Numerical Prediction for Space Storm,
%Harbin Institute of Technology, Shenzhen 51805, People’s Republic
%of China}



\begin{abstract}
Waves are an integral part of the solar atmosphere, and their characteristics (e.g., dominant period, range of periods, power, and phase angle) change on a diverse spatio-temporal scale. It is well-established observationally that the dominant periods of solar oscillations are 5-min and 3-min in the photosphere and chromosphere, respectively. This shows that the wave spectra and their dominant periods evolve between these two layers.  We present observational results that demonstrate variations of the dominant period with heights in the photosphere and chromosphere. Six photospheric absorption lines and one chromospheric line are analyzed by using the IRIS data, and the Doppler velocity time series at seven different atmospheric heights are determined.  The wavelet analysis is applied to these time series, and the resulting spectrum of wave periods and its dominant period are deduced at these 
heights, which gives height variations of the dominant period.  The obtained data shows that the dominant period decreases with height, and that there are also changes in the range of wave periods within the spectrum.  Numerical simulations of filtered wave spectra through the solar atmosphere are also performed, and the obtained results match the observational data.    
\end{abstract}

\section{Introduction} \label{sec:intro}

The most prominent source of non-radiative energy on the Sun is its convection zone, where different waves are generated.  These waves carry their energy 
through the photosphere and dissipate it in the chromosphere and 
corona, leading to the local heating in these regions as well as to their 
atmospheric oscillations (e.g., \citealt{2004suin.book.....S, 2019ASSL..458.....A}).  The existence 
of the photospheric $5$-min oscillations was  
established observationally by \cite{1962ApJ...135..474L}, and explained 
theoretically by \cite{1971ApL.....7..191L}.  Main oscillations of the chromosphere 
are identified with the $3$-min oscillations, with their $2 - 5$-min 
range inside non-magnetic or weak magnetic regions such as 
supergranulation cells (e.g., \citealt{1991mcch.conf....6D, 2002ApJ...567L.165M,2003ApJ...587..806M}).   
However, in the magnetic network, the oscillations range from 
$6$ to $15$-min (e.g., \citealt{1993ApJ...414..345L,2005AN....326..301S}).  

The origin of the oscillations can be understood as a response of the solar atmosphere to different types of waves generated in the convection zone.  If there are atmospheric cavities, the waves
can be trapped in them leading to local oscillations (e.g., \citealt{1998IAUS..185..427D}). 
However, in case there are no such cavities and the waves are propagating,
they can also excite atmospheric oscillations at the local cutoff periods (e.g., \citealt{1991A&A...250..235F, 1993A&A...273..671F,1994A&A...284..976K,1998A&A...337..487S}) and these oscillations decay in time as $t^{-3/2}$.  There are also forced atmospheric oscillations that do not decay in time if the wave source drives them continuously (e.g., \citealt{1995A&A...301..483S, 1999ApJ...519..899H,2003A&A...400.1057M,2003A&A...406..725M, 2024MNRAS.531.4611N}).
Analytical and numerical studies of cutoff frequencies in the solar atmosphere 
were performed for acoustic waves (e.g., \citealt{1998A&A...337..487S,2006PhRvE..73c6612M,2014AN....335.1043R, 2016ApJ...827...37M, 2020ApJ...896L...1M,2024RSPTA.38230218K}), and for magnetic flux tube waves (e.g., \citealt{2004ESASP.547....1R, 2006RSPTA.364..447R, 2006ApJ...640.1153C, 2007ApJ...659..650M,2010ApJ...709.1297R,2013ApJ...763...44R, 2013ApJ...772...90L,2017ApJ...840...26L,2015A&A...577A.126M,2017SoPh..292...31W, 2023LRSP...20....1J,2023A&A...672A.105P}).  
The existence of the wave cutoffs and their variations with the atmospheric height on the Sun has been established observationally by  \cite{2016ApJ...819L..23W, 2018MNRAS.479.5512K}, and \cite{2024ApJ...966..187S} for magnetic-free 
and weak magnetic regions, and by \cite{2009ApJ...692.1211C} and \cite{2018A&A...617A..39F} for solar magnetic regions such as sunspots, sunspot umbra, pores, and faculae.

In theoretical work performed by \cite{2012MNRAS.421..159F,2016Ap&SS.361...23F}, it was shown that a spectrum of waves generated in the convection zone significantly changes its shape when it is transferred through atmospheres of solar-type stars.  The filtered wave spectrum shows fewer frequencies and a more prominent dominant period than the initial broader spectrum.  Their results showed that the dominant period decreases from photospheres to chromospheres of these stars. They also found 
a linear relationship between the frequency of the maximum oscillation amplitude and the acoustic cutoff frequency for such stars, which confirmed the original results obtained by \cite{2010ApJ...713L.182S} and \cite{2011ApJ...743..142H} who used Kepler data. These results demonstrate that the dominant periods of the filtered wave spectrum may be different than the local cutoff frequency at the same atmospheric height.  Oscillations in solar-like stars were observed by the Kepler mission (e.g., \citealt{2011ApJ...732L...5C,2011Sci...332..213C,2011ApJ...743..143H}); however, the predicted variations with atmospheric heights are unlikely to be observed in solar-like stars because the effects are small (e.g., \citealt{2007ApJ...663.1315B}).  On the other hand, such variations may be detected in the solar atmosphere, and this work reports on first direct measurements of dominant wave periods and their variations with height in the solar atmosphere using the spectroscopic observations from Interface Region Imaging Spectrograph (IRIS; \citealt{2014SoPh..289.2733D}).

The paper is organized as follows: Section~\ref{sec:obs} describes the performed observations and analysis of the obtained data.   
The main observational results are presented in Section~\ref{sec:results_obs}, and Section~\ref{sec:results_sim} is dedicated to the numerical simulation and comparison of the obtained numerical results to the data.  Finally, 
the discussion and conclusion are given in Section 5.   
\section{Observation and Data Analysis} \label{sec:obs}
\begin{figure}[ht!]
\includegraphics[trim = 1.0cm 0.0cm 1.0cm 1.5cm, scale=1.1]{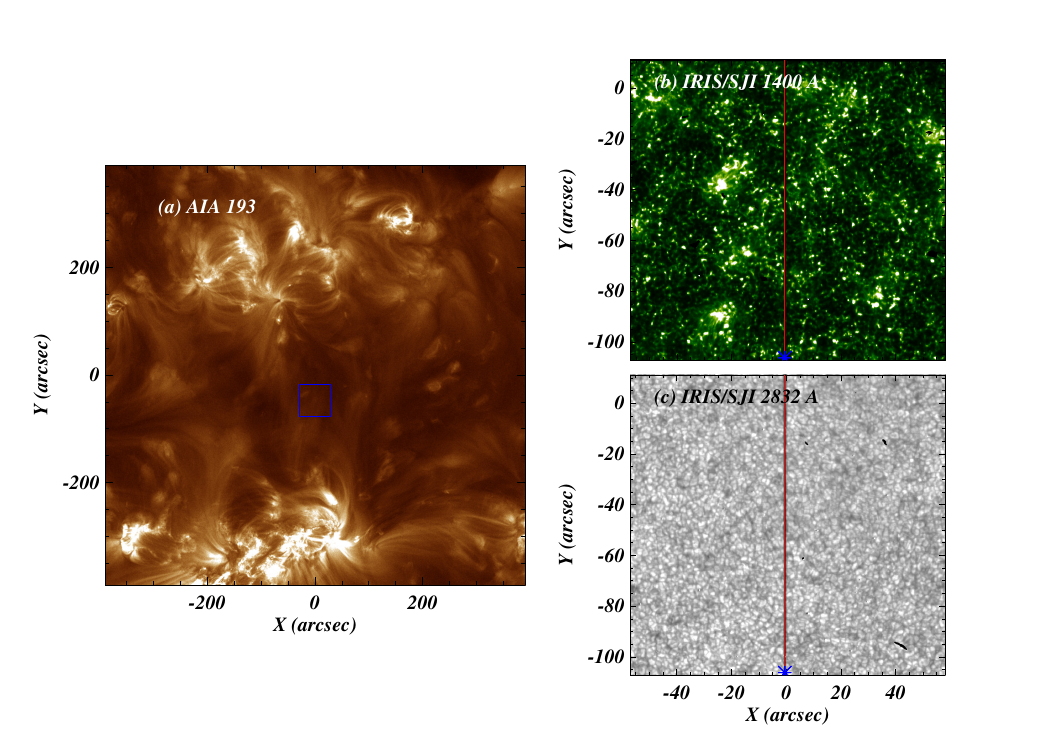}
\caption{The left panel shows the AIA~193~{\AA} map and the overplotted blue box is the region observed by IRIS. The top-right panel shows the transition-region image (IRIS/SJI 1400~{\AA}) corresponding to the blue box, while the bottom-right panel shows the photospheric image (i.e., IRIS/SJI 2832~{\AA}). The overplotted red line in both right panels is the slit location, and IRIS has captured the spectra from the region below the slit.
\label{fig:ref_fig}}
\end{figure}

On November 16$^{th}$, 2013, IRIS captured the far-ultraviolet (1331.7{--}1358.4 and  1389.0{--}1407.0~{\AA}) and near-ultraviolet (2782.7{--}2835.1~{\AA}) spectra in sit-n-stare mode along with the slit-jaw-images (SJI) of quiet Sun (QS). Here, note that QS is located very close to the disk centre, i.e., the central x and y coordinates are close to 0$"$ and -49$"$, respectively. We have utilized images from Atmospheric Imaging Assembly (AIA; \citealt{2012SoPh..275...17L}) and IRIS/SJI to display the observed region (Figure~\ref{fig:ref_fig}). For the alignment, first of all, we have aligned IRIS/SJI~1400~{\AA} with AIA~304~{\AA}. Further, AIA~304~{\AA} is aligned with AIA~193~{\AA}. The panel (a) of Figure~\ref{fig:ref_fig} shows aligned coronal image (i.e., AIA~193~{\AA}). It shows the existence of some active regions (ARs) (see the bright regions above and below the blue rectangular box). We also see the relatively dark area in between the ARs, and this relatively dark area is QS. Note that we have drawn a blue rectangular box in the QS, and the blue rectangular box outlines the spatial region observed by IRIS. The field of view (FOV) of IRIS/SJI is 129$"$$\times$120$"$. Further, we showed the transition-region (panel b) and photospheric view (panel c) of the QS region (corresponding to the blue box) using IRIS/SJI 1400~{\AA} and 2832~{\AA} filters. Here, it should be noted that the vertical red line in both panels on the right show the IRIS slit position. IRIS captured the spectra from the region along the red slit from 07:33:45~UT to 08:08:12~UT. Hence, we have time-series for around 34 minutes. Further, note that the spectra are captured with the exposure time (cadence) of 16.7 s (17 s). The blue asterisk signs in the panels (b) and (c) show a particular spatial location for which wavelet power maps are shown in Figure~\ref{fig:domin_histogram}. Using line-of-sight magnetic field observations from Helioseismic and Magnetic Imager (HMI) onboard Solar Dynamic Observatory (SDO), we investigated the LOS magnetic field along the IRIS slit (i.e., red lines in panels (b) and (c) of Figure~\ref{fig:ref_fig}) from a one particular time (i.e., 07:33:45~UT). The mean magnetic field is 5 G, and it justifies that the observed region is QS.\\
%%%%%%%%%%Figure 1%%%%%%%%%%%%%%%%%%%%%%%%%%%%%%%%%%%%%
%%%%%%%%%%%%%%%%%%%%%%%%%%%%%%%%%%%%%%%%%%%%%%%%%%%%%

In the present work, we have utilized 7 different spectral lines from NUV spectra (see Table~\ref{tab:table1}). The six spectral lines are photospheric absorption lines, and the last spectral line (i.e., Mg~{\sc ii} k 2796.35~{\AA}) is a chromospheric emission line (see Table~\ref{tab:table1}). Actually, the Mg~{\sc ii} k~2796.35~{\AA} is an optically thick line, and mostly, it has two peaks (k2v and k2r) and one dip (k3). These peaks and dips from at different heights in the solar atmosphere, i.e., Mg~{\sc ii} k2r (Mg~{\sc ii} k2r) form at a height of 1.2 Mm (1.55 Mm) while Mg~{\sc ii} k3 forms at a height of 2.2 Mm (\citealt{2013ApJ...772...89L}). Hence, we have nine different heights between the photosphere and chromosphere, and they cover the height from 0.17 Mm to 2.2 Mm.
%These spectral lines form at different heights (i.e., from 0.17 Mm to 1.8 Mm) in the solar atmosphere. %, see the last two columns of Table~\ref{tab:table1}
%%%%%%%%%%%%%%%%%Tabel 1%%%%%%%%%%%%%%%%%%%%%%%%%%%
\begin{table}[]
\caption{The used spectral lines along with their wavelength, formation height, and uncertainties in the formation height.}
    \centering
    \hspace{-2.0cm}
    \begin{tabular}{|c|c|c|c|c|}
       \hline
Sr. No. & Spectral Line & Wavelength (\AA) & Formation Height (Mm) & Uncertainty (Mm) \\
\hline
1 & Ni~{\sc i} & 2815.179 & 0.17 & 0.03 \\ 
\hline
2 & Fe~{\sc i} & 2792.327 & 0.38 & 0.03 \\
\hline
3 & Fe~{\sc i} & 2793.223 & 0.50 & 0.04 \\
\hline
%4 & Cr~{\sc ii} & 2801.584 & 0.58 & 0.03 \\
4 & Ni~{\sc i}  & 2799.347 & 0.68 & 0.07 \\
\hline
5 & Fe~{\sc i}  & 2814.114 & 0.76 & 0.11\\
\hline
6 & Mn~{\sc i}  & 2801.907 & 0.83 & 0.10\\
\hline
7 & Mg~{\sc ii} k2v  & 2796.35 & 1.20 & ..\\
\hline
8 & Mg~{\sc ii} k2r & 2796.35 & 1.55 & ..\\
\hline
7 & Mg~{\sc ii} k3 & 2796.35 & 2.20 & ..\\
\hline
    \end{tabular}
    \label{tab:table1}
\end{table}
%%%%%%%%%%%%%%%%%%%%%%%%%%%%%%%%%%%%%%%%%%%%%%%%%
All the necessary details (e.g., ion, wavelength, formation height, and uncertainty in the formation height) of the used lines are described in Table~\ref{tab:table1}. 
Note that all the details about the spectral lines mentioned in Table~\ref{tab:table1} are taken from the ITN39\footnote[1]{https://iris.lmsal.com/itn39/photospheric.html \label{fn_1}}. The formation height of Mg~{\sc ii} k 2796.35~{\AA} line varies from 0.6 Mm (k1 peak) to 2.2 Mm (central k3 peak; \citealt{1981ApJS...45..635V, 2013ApJ...772...89L}). In the present work, we have used only Mg~{\sc ii} k2r, Mg~{\sc ii} k2v, and Mg~{\sc ii} k3.\\

We have followed the methodology described in ITN 39$^{\ref{fn_1}}$ to fit the photospheric absorption lines (i.e., the first six lines in Table~\ref{tab:table1}). The Gaussian fit allows us to estimate the intensity, centroids, and sigma at each spatial and temporal location. Then, the centroids are converted into the Doppler velocities with the help of rest wavelengths. Mostly, the Mg~{\sc ii line profiles} have two peaks (i.e., k2v and k2r) and a central dip region (i.e., k3). Therefore, it must be noted that the single Gaussian fit is not appropriate for Mg~{\sc ii} as it is an optically thick line (\citealt{2013ApJ...772...90L}). We applied a method described by \citealt{2013ApJ...778..143P} (i.e., using an inbuilt routine iris$\_$get$\_$mg$\_$features$\_$lev2.pro) to get the Doppler velocities of Mg~{\sc ii} k2r, Mg~{\sc ii} 2v, and Mg~{\sc ii} k3. The algorithm used in this iris$\_$get$\_$mg$\_$features$\_$lev2.pro is summarized below. They have employed an extremum-finding algorithm within -40 km/s to 40 km/s (i.e., within the small spectral region) to wavelengths of all maxima and minima. Most of the profiles have two maxima and one minima, and in this case, the middle minima is Mg~{\sc ii} k3. If minima are more than one, then the minima with the lowest intensity are considered as Mg~{\sc ii} k3. Lastly, if no minima are found, then a parabola is fitted within the range of 5 km/s of Mg~{\sc ii} k to get the line centre wavelength. After performing all these steps, we have constructed Doppler velocity-time series (hereafter DTS) from all the spatial locations along the slit and for all spectral lines.\\  

%The estimation of rest wavelength is very crucial, and usually, we use cool photospheric lines (i.e., singly and doubly ionized lines) to estimate the rest wavelength of relatively hotter lines. However, in the present study, all the photospheric lines are cool, therefore, we have used their standard wavelength (i.e., the wavelength mentioned in column 3 of table~\ref{tab:table1}). Here, we mention that the last spectral line (i.e., Mg~{\sc ii} k 2796.35~{\AA}) is the chromospheric, not the photospheric. We need to estimate the rest wavelength of this line, and we can not use the standard wavelength. Hence, we estimated the rest wavelength for this Mg~{\sc ii} k. Finally, using this rest wavelength, we have estimated the Doppler velocity for all spectral lines. Finally, we used to Doppler velocity time series to perform the wavelet analysis. 

\section{Observational Results} \label{sec:results_obs}
\begin{figure}[ht!]
\includegraphics[]{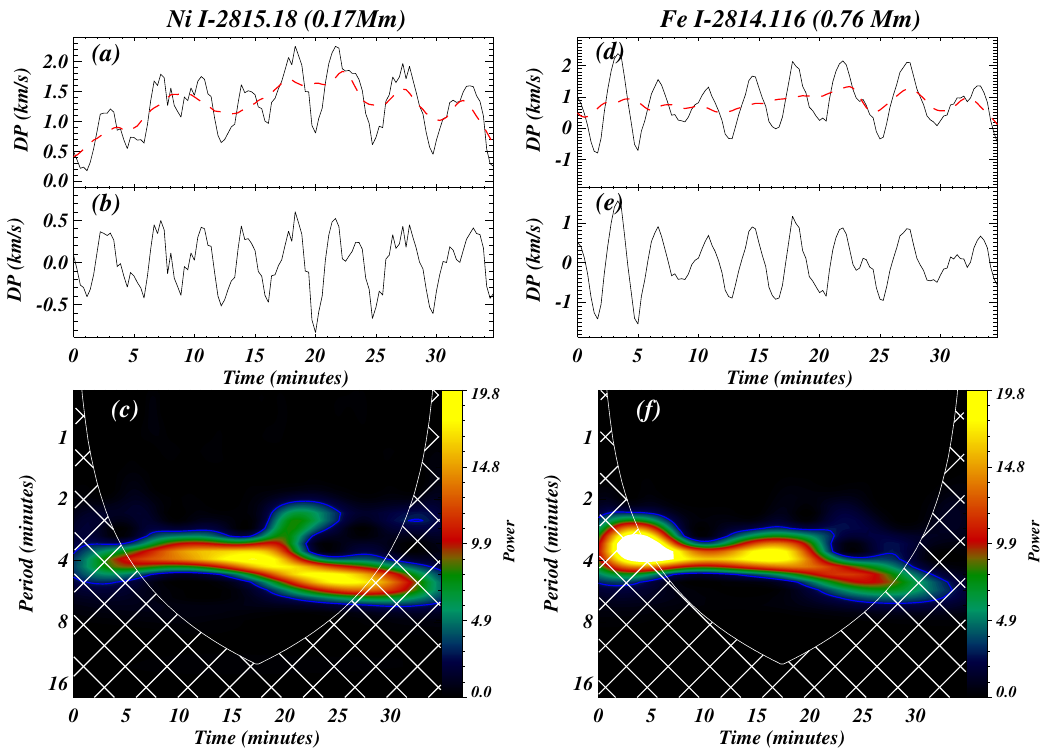}
\caption{The panel (a) displays the original DTS (black curve) and smoothed DTS (red dashed curve) from one particular spatial location in Ni~{\sc i} 2815.18~{\AA}. The smoothed DTS is produced using gauss$\_$smooth.pro with a window size of 12 points. The original DTS-smoothed DTS (black curve-red curve) is displayed in panel (b). Finally, the wavelet power map is shown in panel (c). The overplotted blue contour is the 95\% significance level, and the white cross-hatched area outlines the cone-of-influence. The panels (d), (e), and (f) have the same information but for Fe~{\sc i} 2814.116~{\AA}.  
\label{fig:wavelet_power}}
\end{figure}

Panel (a) of Figure~\ref{fig:wavelet_power} shows the DTS of Ni~{\sc i} 2815.18~{\AA} from one particular spatial location, and the location is marked by a blue asterisk in the panels (b) and (c) of Figure~\ref{fig:ref_fig}. The overplotted red-dashed line is the smoothed DTS using a window of 12 points (i.e., the total time corresponds to 12 points; $t = 17 s \times 12 = 204$ s). The smoothed DTS (red curve) is subtracted from the original DTS (black curve) to get the detrended DTS, and it is displayed in panel (b). We have applied wavelet analysis (\citealt{1998BAMS...79...61T}) on the detrended DTS to deduce the power map, and the deduced power map is shown in Figure~\ref{fig:wavelet_power}(c). The blue contours outline the significant power above 95\% confidence levels. The white cross-hatched area in panel (c) represents the cone of influence (COI). The power is significant around 4 minutes from time $t = 0$ to around $t= 20 $ minutes. After $t = 20$ minutes, the dominant periods shift towards the longer period (i.e., around 6 minutes). Here, please note that \cite{2003ApJ...595L..63D} have reported a similar transition from shorter to longer periods but in the active-region plage.

\begin{figure}[ht!]
\includegraphics[trim = 0.0cm 2.0cm 0.0cm 2.1cm,scale=1.0]{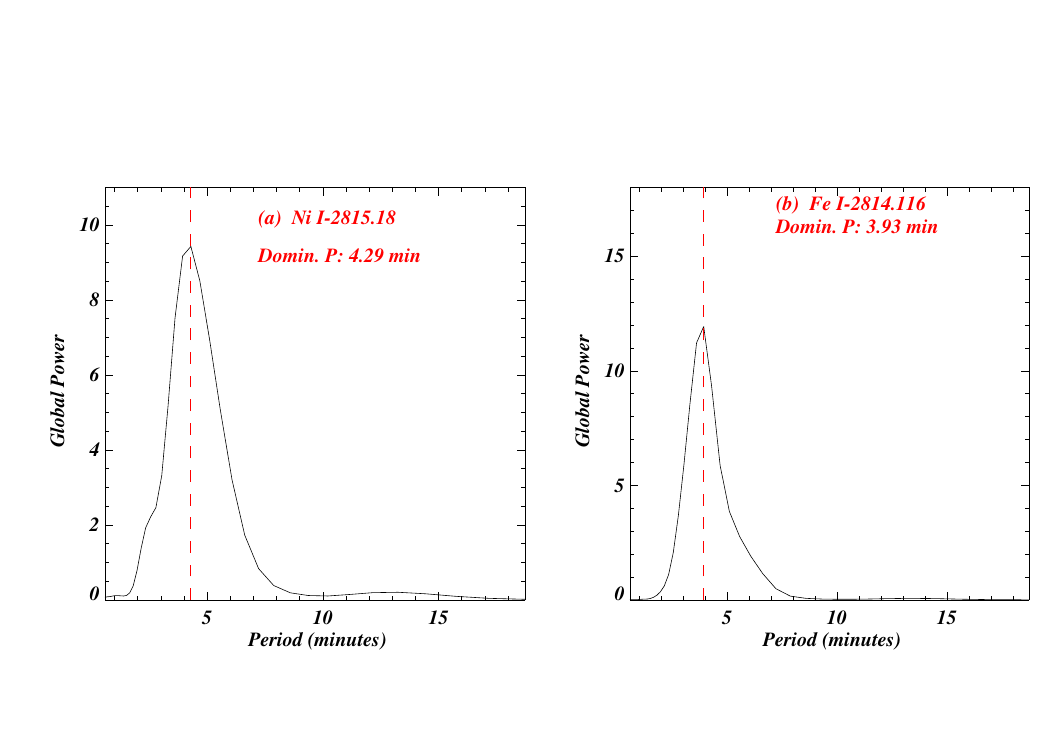}
\includegraphics[trim = 0.0cm 1.0cm 0.0cm 2.1cm,scale=1.0]{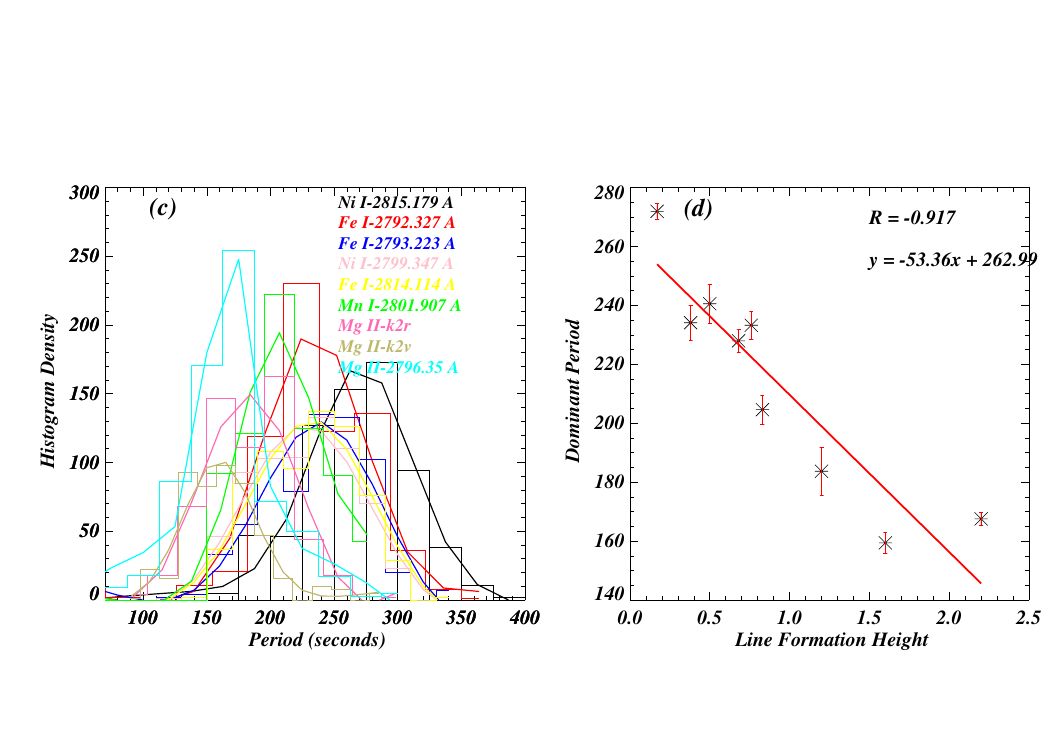}
\caption{The global wavelet power deduced using wavelet power maps shown in panels (c) and (f) in Figure~\ref{fig:wavelet_power} are displayed in panels (a) and (b), respectively. The vertical red-dashed lines are the dominant period from this particular location. We calculated the dominant period from each location and for each spectral line to produce the histogram, and such histograms are displayed in panel (c) for all lines. Finally, applying Gaussian fit on each histogram deduces the mean period for each spectral line. The correlation between the mean period and formation height is shown in panel (d).
\label{fig:domin_histogram}}
\end{figure}

In the same fashion, the original DTS with smoothed time series (panel d), detrended time series (panel e), and wavelet power map (panel f) are displayed from the same spatial location but from different line (i.e., Fe{\sc i} 2814.116~{\AA} line) in Figure~\ref{fig:wavelet_power}. The behavior of the power in this line is the same as in Ni~{\sc i} 2815.18~{\AA}. Although most importantly, on average, the periods are shorter in the Fe~{\sc i} 2814.116~{\AA} than in Ni~{\sc i} 2815.18~{\AA}.
The global power vs. period corresponding to wavelet power maps shown in Figures~\ref{fig:wavelet_power}(c) and~\ref{fig:wavelet_power}(f) are displayed in panels (a) and (b) of Figure~\ref{fig:domin_histogram}, respectively. The dominant periods in Ni~{\sc i} 2815.18~{\AA} and Fe~{\sc i} 2814.116~{\AA} are 4.29 and 3.93 minutes (see red dashed lines in panels (a) and (b)).

%%%%%%%%%%Figure 2%%%%%%%%%%%%%%%%%%%%%%%%%%%%%%%%%%%%%
%%%%%%%%%%%%%%%%%%%%%%%%%%%%%%%%%%%%%%%%%%%%%%%%%%%%%
%
%%%%%%%%%%Figure 3%%%%%%%%%%%%%%%%%%%%%%%%%%%%%%%%%%%
%%%%%%%%%%%%%%%%%%%%%%%%%%%%%%%%%%%%%%%%%%%%%%%%%%%%%

Using this method, the dominant periods are calculated for all 700 locations along the slit and produced the histogram of the dominant period for all 7 lines (see Table~\ref{tab:table1}), and these histograms are shown in the Figure~\ref{fig:domin_histogram}(c). Further, each histogram is fitted with the Gaussian function and the fitted Gaussian curve is overplotted on the corresponding histogram using the same color, for instance, black histogram and black solid line (Gaussian fit) for Ni~{\sc i} 2815.18~{\AA}. The fitted Gaussian function gives the mean period and standard deviation ($\sigma$) of all spectral lines, i.e., the mean period at all formation heights mentioned in Table~\ref{tab:table1}). The formation height, mean dominant periods, and the $\sigma$ of each spectral line are listed in Table~\ref{tab:table2}.
%%%%%%%%%%%%%%%%%Tabel 1%%%%%%%%%%%%%%%%%%%%%%%%%%%
\begin{table}[]
\caption{The mean dominant period, $\sigma$, and formation height of each spectral line. The mean dominant period and $\sigma$ are deduced from the Gaussian fit on the histogram (cf., Figure~\ref{fig:domin_histogram}).}
    \centering
    \hspace{-2.0cm}
    \begin{tabular}{|c|c|c|c|c|}
       \hline
Sr. No. & Spectral Line & Formation Height (Mm) 
 & Dominant Period (s)  & Gaussian Sigma (s) \\
\hline
1 & Ni~{\sc i} & 0.17 & 271.93 & 38.83 \\ 
\hline
2 & Fe~{\sc i} & 0.38 & 234.10 & 40.57 \\
\hline
3 & Fe~{\sc i} & 0.50 & 240.50 & 49.19 \\
\hline
4 & Ni~{\sc i} & 0.68 & 227.90 & 53.80 \\
\hline
5 & Fe~{\sc i} & 0.76 & 233.20 & 50.27 \\
\hline
6 & Mn~{\sc i} & 0.83 & 204.60 & 30.38 \\
\hline
7 & Mg~{\sc ii} k2r & 1.2 & 183.69 & 45.17 \\
\hline
8 & Mg~{\sc ii} k2v & 1.55 & 159.408 & 28.35 \\
\hline
9 & Mg~{\sc ii} k3 & 1.80 & 167.0 & 23.69 \\
\hline
    \end{tabular}
    \label{tab:table2}
\end{table}
%%%%%%%%%%%%%%%%%%%%%%%%%%%%%%%%%%%%%%%%%%%%%%%%%
On average, the mean dominant period decreases with the increasing value of formation height, see Table~\ref{tab:table2}. Lastly, the correlation between formation height and mean dominant period is displayed in Figure~\ref{fig:domin_histogram}(d). The solid red line is a fitted straight line on the mean dominant period vs formation height which clearly shows that the mean dominant period falls linearly with the increasing height. The linear fit has very high Pearson's coefficients (i.e., 0.944; Figure~\ref{fig:domin_histogram}(d)). 

\section{Numerical Simulations} \label{sec:results_sim}
The two and half-dimensional (2.5-D) numerical simulations of the solar atmosphere are performed with the JOANNA code 
\citep{Wojciketal2019b,Wojciketal2019c,Wojciketal2019a}. 
This code solves the non-ideal and non-adiabatic 
two-fluid equations with ionization/recombination 
terms within a box that for these simulations 
is specified along horizontal ($x$-) and vertical ($y$-) directions as 
$(-2.56\le x\le 2.56)$$\times$ $(-2\le y\le 20)$\,Mm$^{2}$. 
The system is assumed invariant along $z$-direction 
($\partial/\partial z=0$) but the transversal ($z$-) components 
of ion and neutral velocities ($V_{\rm i,n\, z}$) 
and magnetic field ($B_{\rm z}$) 
are 
%not identically zero. 
allowed to vary. 
Below the altitude $y=3.12$\,Mm, 
a uniform grid of $256\times 256$ cells with cell's size 
$\Delta x=\Delta y=20$\,km 
is set, 
while higher up the grid is stretched along $y-$direction, 
dividing 
it into $96$ cells whose size steadily grows with height. 
At $y=-2$\, Mm and $y=20$\, Mm all plasma quantities are 
fixed in time to their magnetostatic values. 
The only exception is the bottom boundary at which 
vertical components of ion and neutral velocities, 
$V_{\rm i,n\, y}$, are set to $0.2$\, km\,s$^{-1}$. 
%and the transversal component of magnetic field, $B_{\rm z}$, 
%is assumed to be $2$\, G. 
%The latter values mimic the emerging at the depth $y=-5$\, Mm 
%magnetic field. 
%%results from the requirement of balancing 
%%outgoing mass flux by its inflowing counterpart. 
%%boundary conditions are implied, 
%%while at the bottom inflow boundary conditions are used with 
%%all plasma quantities fixed at all time $t\ge 0$\, s 
%%to their magnetostatic values. 
%%with 
%
The stretched grid implemented in the top zone results in 
damping of the incoming signal, leading to negligibly small 
numerically-induced reflections from the top boundary. 
Left and right boundaries are set to be periodic. 
Note that an assumed constant gravitational acceleration 
points towards 
negative $y-$axis and it is given as 
$\mathbf{g} = [0, -g, 0]$ with $g = 274.78$\,m\,s$^{-2}$; 
for more details, see \cite{2022Ap&SS.367..111M}. 

\begin{figure}[!ht]
\centering

\includegraphics[trim = 5cm 1.1cm 10cm 3cm, scale=0.30]{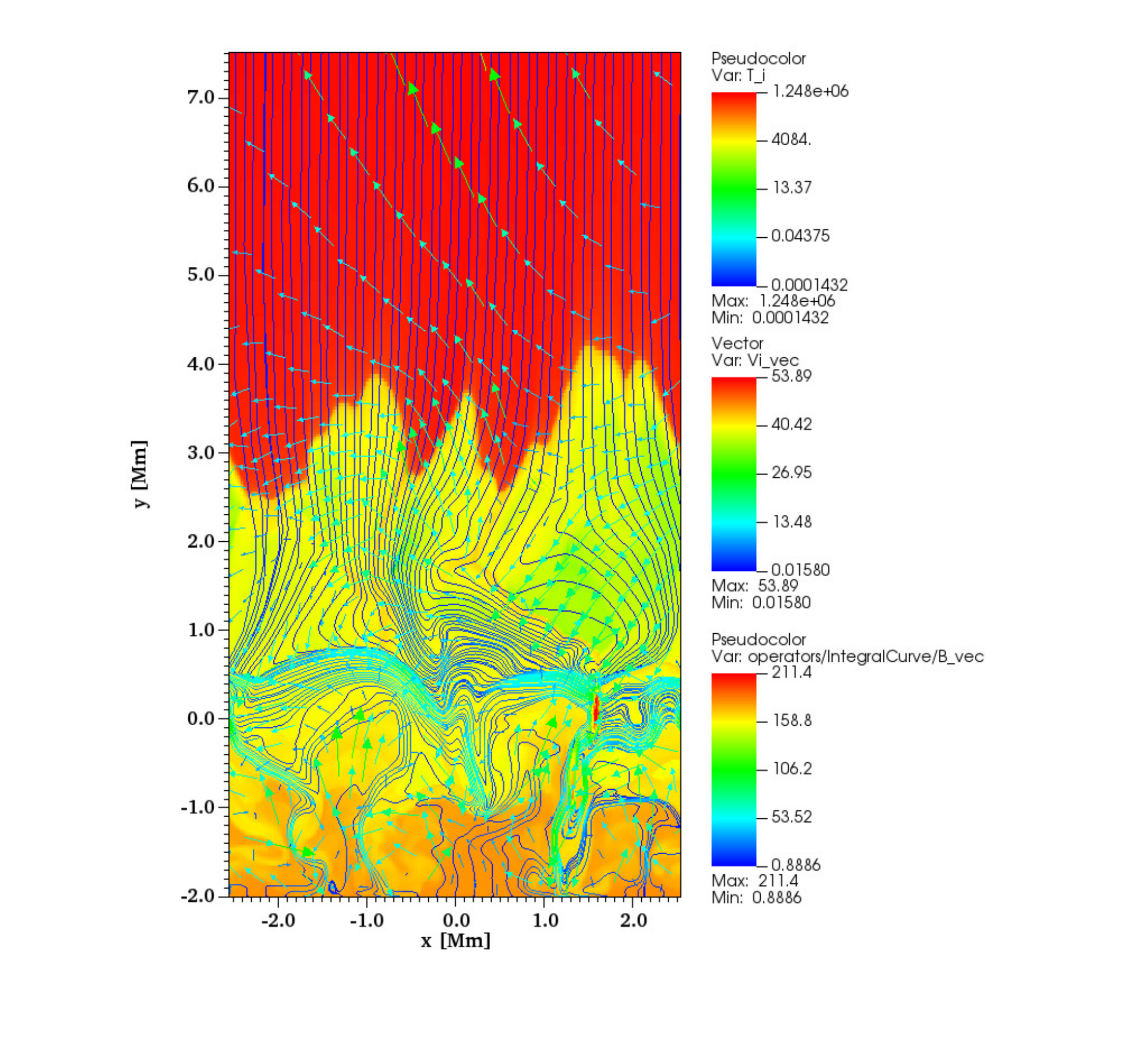}

\includegraphics[bb=0 0 1200 900, trim = 10cm 1cm 10cm 1.5cm,scale=0.60]{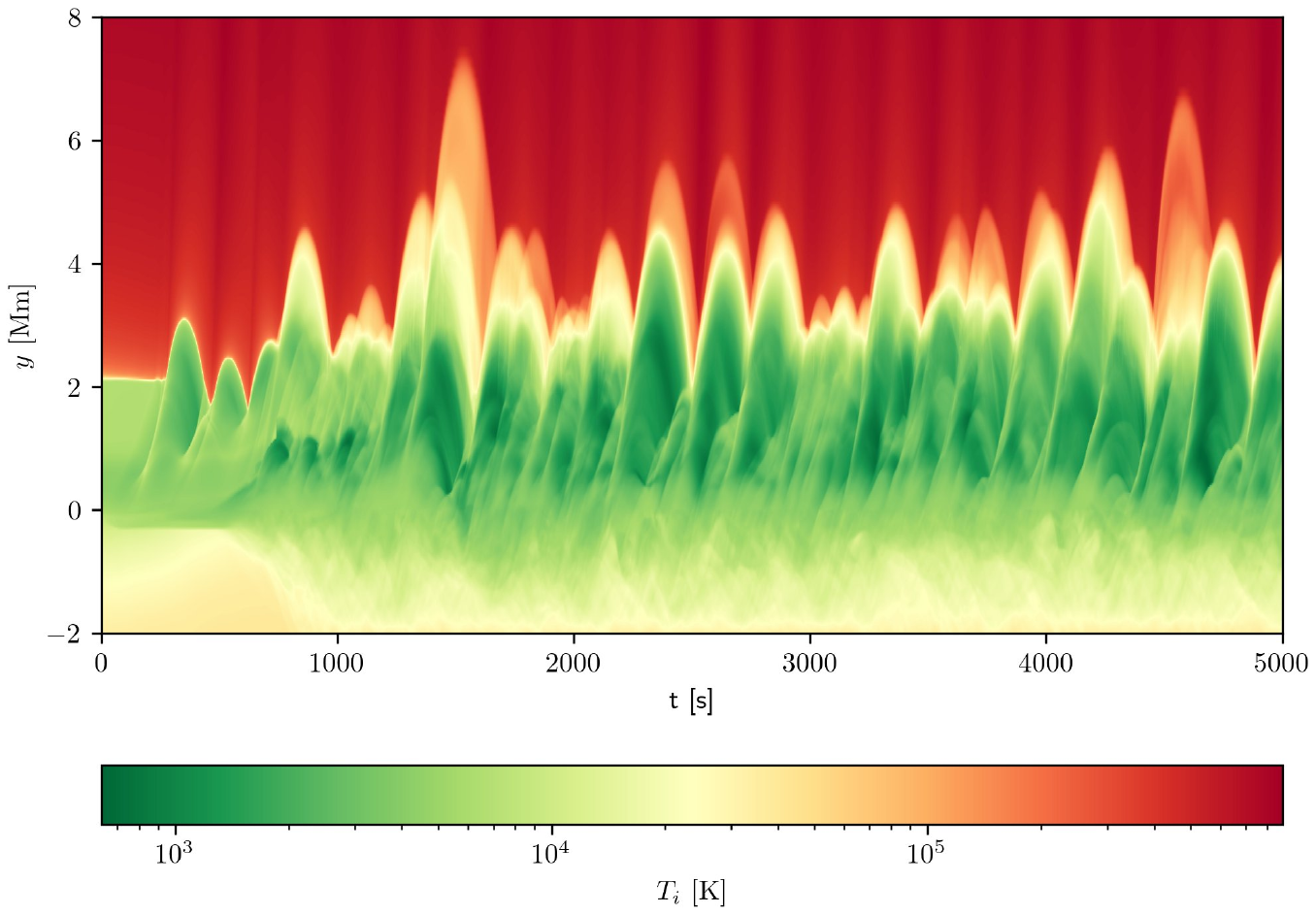}\\
\caption{The figure shows the spatial profile of $\log{T_{\rm i}(x,y,t=5\cdot 10^3\, {\rm s}})$, overlaid by magnetic field lines and $ V_{\rm i}$ vectors (top panel). The time-distance plots for horizontally averaged $T_{\rm i}$ is displayed in the bottom panel.}
\label{fig:sim1}
\end{figure}

Our simulations are initiated at $t=0$\, s by implementing a hydrostatic solar atmosphere supplemented by the Saha equation 
\citep{1920Natur.105..232S} 
with the semi-empirical temperature, $T(y)$, model of \cite{AvrettLoeser2008}. 
%See Fig.~\ref{fig:T-n_init}, left. 
This temperature, 
which 
%initially 
%(
at $t=0$\, s
%) 
is identical for ions and neutrals, 
$T_{\rm i}=T_{\rm n}=T(y)$, 
%uniquely 
determines equilibrium 
ion and neutral mass densities 
%(Fig.~\ref{fig:T-n_init}, right) 
and gas pressures \citep{2020ApJ...896L...1M}. 
%Convective instabilities are seeded by launching initially 
%%(at $t=0$\, s) 
%a 
%small ($1$\, m\,s$^{-1}$) 
%and random 
%signal in 
%vertical components of 
%ion and neutral velocities. 
%These instabilities 
Convective instabilities 
self-generate from a small, initial, random signal and 
self-evolve in time, being 
most prominent below the photosphere.  
They lead to  
turbulent fields that mimic the convection with granulation 
cells at its top (Fig.~\ref{fig:sim1}, top panel). 
Such turbulent fields reshape the 
%initial, 
%uniform and unidirectional 
magnetic field which is 
%initially (
at $t=0$\, s 
%) 
taken in the form of 
%a magnetic carpet being overlaid by the straight magnetic field given as
%
%\begin{equation}
$\textbf{B} = 
%B_{\rm a} \left[\cos \left(\frac{x+L_{\rm B}}{\Lambda_B}\right), 
%-\sin \left(\frac{x+L_{\rm B}}{\Lambda_B}\right), 0\right] e^{-y/\Lambda_B} + 
\left[0, B_y, B_z\right]\, 
$. 
%    \label{eq:magnetic}
%\end{equation}
%
Here, 
%$B_{\rm a}=2.5$ G, 
$B_{\rm y}=5$ G and $B_{\rm z}=0.5$ G correspond respectively to 
%a magnetic carpet (Niedziela et al. 2024, in press), 
%modelled by the set of arcades, 
vertical and transversal components. 
%of magnetic field. 
%, 
%$L_{\rm B} = 1.88$\, Mm is the half-size of a single arcade, and 
%$\Lambda_B = 2 L_{\rm B} / \pi$ denotes a height 
%over which $B$ falls off $e-$times. 
%
%%$[B_{\rm x},B_{\rm y},B_{\rm z}]=[0,5,2]$\, Gs 
%%[\delta B_{\rm x},10,2]$\, Gs 
%%the magnitude of $10$\,G 
The reshaped field 
%into well developed 
contains 
complex structures with magnetic flux-tubes 
%below the transition region 
(Fig.\, \ref{fig:sim1}, top panel). 
%Here $\delta B_{\rm x}=0$ for $y>0$ Mm and $50$ Gs in 
%the convection zone that is located for $y\le 0$ Mm. 
%This horizontal component of magnetic field is initially implemented 
%to sustain the average magnetic field in the convection zone. 
The spatial profile of $\log\,T_{\rm i}(x,y)$ exhibits the perturbed pattern with well seen jets at the transition region (Fig.\, \ref{fig:sim1}, top panel), which initially was located at $y=2.1$\, Mm.
%%%%%%%%%%%%%%%%%%%%%%%%%%%%%%%%%%%%%%%%%%%
%%%%%%%%%%%%%%%%%%%%%%%%%%%%%%%%%%%%%%%%%%%

\begin{figure}[!ht]
\centering

\includegraphics[trim = 10cm 0cm 10cm 1cm,scale=0.60]{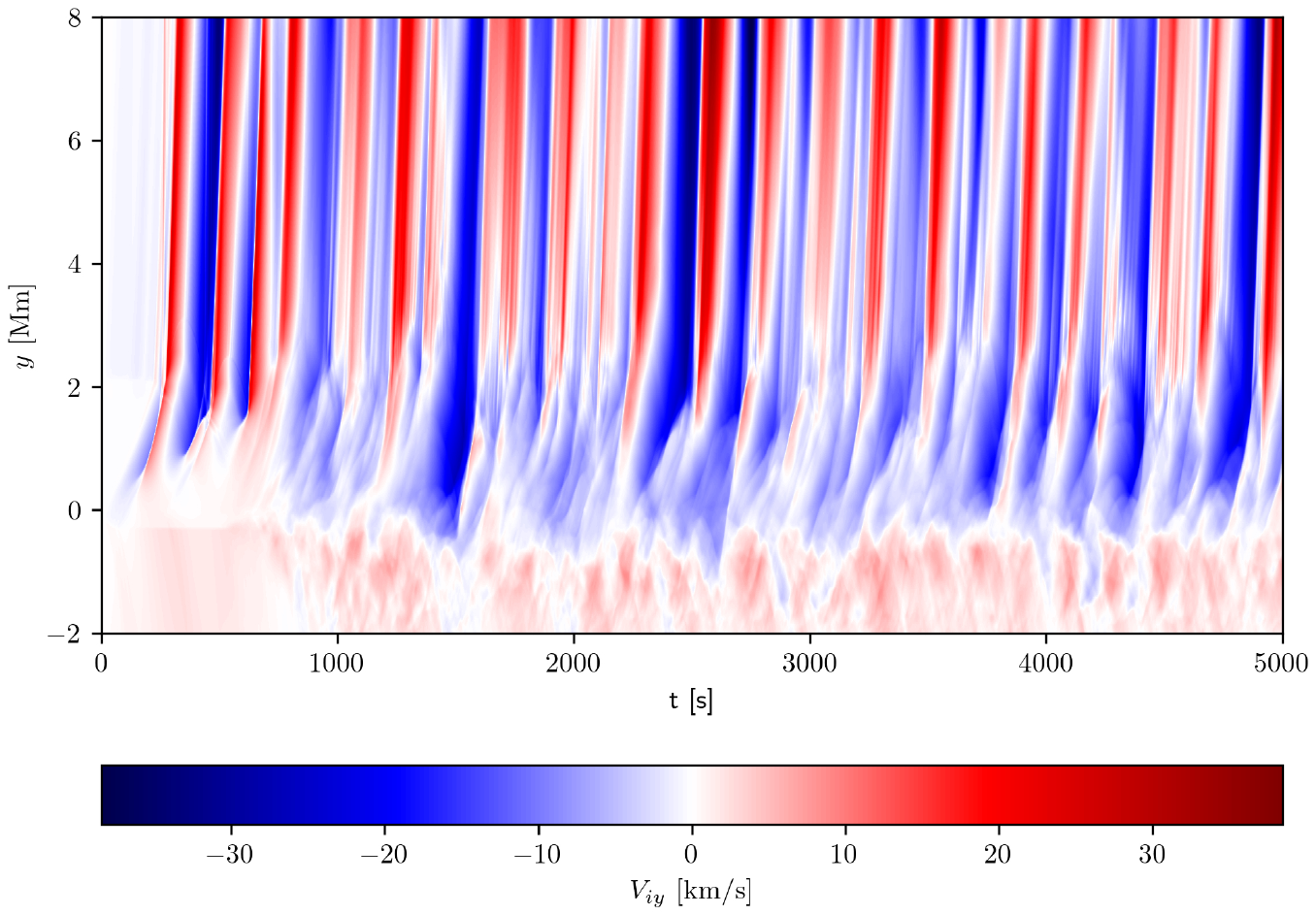}

\includegraphics[trim = 10cm 2.0cm 10cm 2.5cm,scale=0.68]{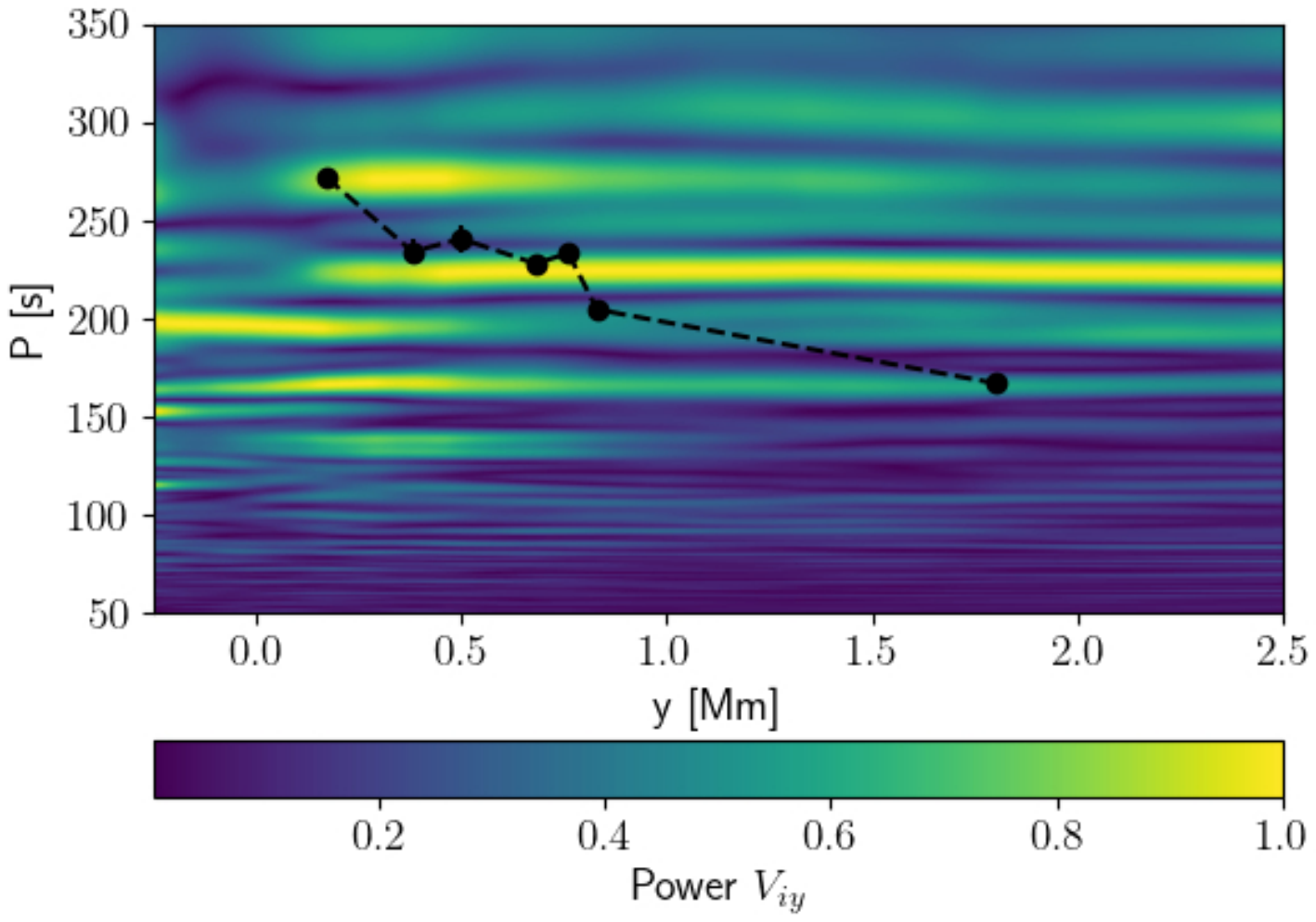}
%\hspace{0.5cm}
\caption{Time-distance plots for horizontally $V_{\rm iy}$ is shown in the top panel. Fourier power spectrum of dominant wave period $P$ for 
%horizontally averaged 
%vertical component of ion velocity, 
$V_{\rm iy}$ 
taken from bottom-left panel for 
$2\cdot 10^3\, {\rm s} \le t \le 5\cdot 10^3\, {\rm s}$. 
%vs. height. 
The dots correspond to the observational data presented in this paper.}
\label{fig:sim2}
\end{figure}
%%%%%%%%%%%%%%%%%%%%%%%%%%%%%%%%%%%%%%%%%%%%%%

Figure\, \ref{fig:sim1} (top panel) shows 
the horizontally averaged $T_{\rm i}$. The self-generated granulation expels the fluids from the lower atmospheric layers into the transition region and higher up into the low corona with the largest jets arriving to $y\approx 8\; \rm Mm$ at $t\approx 1700$\, s (bottom panel; Figure~\ref{fig:sim1}). The existence of such jets (spicules) in the chromosphere has been well-established by the following observations (e.g., \citealt{2014ApJ...792L..15P, 2015ApJ...806..170S}).  Consequently, the transition region experiences oscillations, and ion flows reach their maximum velocities of about $54$\; km s$^{-1}$ (top-left) with the vertical component of ion velocity, $V_{\rm iy}$, reaching about $35$\; km s$^{-1}$ (bottom-left).

Figure\, \ref{fig:sim2} (bottom panel) shows the wave period, $P$, of the excited two-fluid waves that is computed from the Fourier power spectra, which are obtained from the results presented in the top panel of this figure for $2\cdot 10^3\, {\rm s} \le t \le 5\cdot 10^3\, {\rm s}$. 
A few dominant periods can be distinguished, namely, $P \approx 300$\, s 
(observed for $y<0.5$\, Mm) and $200\, s < P < 250$\, s, whose presence is 
illustrated by the yellow strips in this figure.
The period of $300$\, s is absent in the chromosphere as such long period waves are 
evanescent and they correspond to non-propagating waves, which is consistent with 
the results shown in Figure 4.
%\citep{kuzma2024RSPTA.38230218K}. 
%As a result, we infer that the plasma background 
%is altered in time, increasing the cutoff period and 
%allowing so large period waves to propagate 
%from the photosphere through the chromosphere into 
%the corona. 
Waves of the shorter periods 
%of $P \approx 220$\, s 
are seen in the whole atmosphere, but particularly 
in the chromosphere 
%to $y\approx 1$\, Mm 
and 
in the low corona above $y\approx 2.1$\, Mm, 
%and higher up, 
which evidences that such period waves 
propagate freely into the corona. 
Besides these two major wave periods, shorter periods 
waves with $P$ being within the range of 
about $100-150$\, s are also 
%generated by the granulation 
%and they are 
seen 
in the chromosphere and low corona. 
%throughout the whole atmosphere. 
%Some numerically obtained dominant wave periods 
%data 
The
Fourier power spectrum is 
%are close 
similar 
to 
%the new observational data (dots) and 
that recently reported 
by \cite{2024RSPTA.38230218K}, and its  
%and illustrated by diamonds. 
comparison to the 
%observational data of \cite{2016ApJ...819L..23W} (diamonds) and the 
observational data reported in this paper 
%\cite{2018MNRAS.479.5512K} 
(dots) 
reveals good agreement at those atmospheric heights at which 
the observations were performed.\\

%%%%%%%%%%Figure 6%%%%%%%%%%%%%%%%%%%%%%%%%%%%%%%%%%%%%
%\begin{figure}[ht!]
%\includegraphics[]{Map_P_vs_y_Vx2_i.png}
%\caption{ 
%\label{fig:general}}
%\end{figure}
%%%%%%%%%%%%%%%%%%%%%%%%%%%%%%%%%%%%%%%%%%%%%%%%%%%%%
%
%

%\newpage

%\section{Concluding Remarks} \label{sec:discussion}
\section{Summary and Conclusion} \label{sec:discussion}

The main goal of this paper is to establish observationally how wave spectra responsible for the solar oscillations are filtered in the solar atmosphere, specifically, how their dominant periods vary with atmospheric heights.  This paper presents new observational data that is used to determine changes of the dominant wave periods in the solar photosphere and chromosphere.  Observations of six photospheric absorption lines and one chromospheric spectral line located between the atmospheric heights $0.17$ Mm and $1.80$ Mm in the solar atmosphere were performed by the IRIS Observatory, and analyzed here to obtain the data presented and discussed in this paper.  The obtained observational results demonstrate that the dominant periods in the wave spectra that are filtered through these atmospheric layers decrease with heights, and that the observed decrease ranges from the wave period 272 s to 167 s in the photosphere and chromosphere, respectively. 

The observational data is supplemented by numerical simulations of the propagation of a wave spectrum through the solar atmosphere, and comparison of the data to the numerical results is made.  In the performed simulations, the non-ideal and non-adiabatic two-fluid equations for protons + electrons and hydrogen atoms with ionization and recombination included are solved in the presence of a magnetic field with its vertical and transverse strength of 5 G and 0.5 G, respectively.  The obtained numerical results reveal the dominant period of the
filtered wave spectrum, and its variations with height, which are directly compared to the observational data. A good agreement was found between the numerical results and the data (see the bottom panel of Figure~\ref{fig:sim2}), which validates the significance of the physical processes considered in these simulations in the solar atmosphere.

\begin{acknowledgments}
The authors express their thanks to the unknown referee for his/her comments. KM's work was carried out as part of the Space Weather Awareness Training Network (SWATNet) project, funded by the European Union's Horizon 2020 research and innovation program under grant agreement No 955620. Here we would like to thank the IRIS observatory which provides observations of multiple photospheric absorption lines, these lines form at different heights within the solar photosphere. IRIS is a NASA small explorer mission developed and operated by LMSAL with mission operations executed at NASA Ames Research Center and major contributions to downlink communications funded by ESA and the Norwegian Space Centre. %KM's work was done within the framework of the project from the Polish Science Center (NCN) Grant No. 2020/37/B/ST9/00184. 
\end{acknowledgments}
%

%\appendix

%\section{Appendix information}

%\section{Author publication charges} \label{sec:pubcharge}

%\section{Rotating tables} \label{sec:rotate}

%\bibliography{sample631}{}
%\bibliographystyle{aasjournal}

{}



\end{document}